# Matching with Externalities for Context-Aware User-Cell Association in Small Cell Networks

Francesco Pantisano[1], Mehdi Bennis[2], Walid Saad[3], Stefan Valentin[4], and Mérouane Debbah[5]

[1]JRC - Joint Research Centre - European Commission, Ispra, Italy, email: francesco.pantisano@jrc.ec.europa.eu
[2]CWC - Centre for Wireless Communications, Oulu, Finland, email: bennis@ee.oulu.fi
[3]Electrical and Computer Engineering Department, University of Miami, Coral Gables, FL, USA, email: walid@miami.edu
[4] Bell Labs, Alcatel-Lucent, Stuttgart, Germany, email: stefan.valentin@alcatel-lucent.com
[5] Alcatel-Lucent Chair in Flexible Radio, SUPÉLEC, Gif-sur-Yvette, France, email: merouane.debbah@supelec.fr

*Abstract*—In this paper, we propose a novel user-cell association approach for wireless small cell networks that exploits previously unexplored context information extracted from users' devices, i.e., user equipments (UEs). Beyond characterizing precise quality of service (QoS) requirements that accurately reflect the UEs' application usage, our proposed cell association approach accounts for the devices' hardware type (e.g., smartphone, tablet, laptop). This approach has the practical benefit of enabling the small cells to make better informed cell association decisions that handle practical device-specific QoS characteristics. We formulate the problem as a matching game between small cell base stations (SBSs) and UEs. In this game, the SBSs and UEs rank one another based on well-designed utility functions that capture composite QoS requirements, extracted from the context features (i.e., application in use, hardware type). We show that the preferences used by the nodes to rank one another are interdependent and influenced by the existing network-wide matching. Due to this unique feature of the preferences, we show that the proposed game can be classified as a many-to-one matching game with *externalities*. To solve this game, we propose a distributed algorithm that enables the players (i.e., UEs and SBSs) to self-organize into a stable matching that guarantees the required applications' QoS. Simulation results show that the proposed context-aware cell association scheme yields significant gains, reaching up to $52\%$ improvement compared to baseline context-unaware approaches.

## I. Introduction

The deployment of small cell base stations (SBSs) (i.e., picocell, microcell, femtocell), overlaid on current cellular architecture, has emerged as a key solution for meeting the stringent quality-of-service (QoS) requirements of emerging wireless and mobile services. However, reaping the benefits of small cell deployments is contingent upon addressing a number of fundamental challenges that include network modeling and analysis and resource management, among others [1].

In this respect, most existing works on small cell networks have focused on physical layer aspects such as resource management [1–3], user admission control [4], and coordination [5], [6]. In particular, one fundamental challenge in small cell networks is the problem of associating the user equipments (UEs) to their serving SBSs for downlink transmission [1]. For instance, the challenges of the user association problem in small cell networks is significantly different from those dealt with in traditional macro-cellular networks due to the density of SBSs, their heterogeneity (disparate coverage areas and cell sizes), and their limited available resources [1], [2], [4–6]. These unique features of small

This work was supported by the Finnish funding agency for technology and innovation, the U.S. National Science Foundation (Grant CNS-1253731), Nokia Siemens Networks, the Nokia foundation and the Tauno Tonning foundation.

cell networks limit the applicability of macrocell-oriented user association techniques such as [7–10] (and references therein), which can often lead to unbalanced traffic loads at the SBSs and are unable to meet each individual user's QoS.

To this end, one promising approach for addressing the cell association problem is by making the network better informed of its environment and users, hence enhancing its ability to make efficient user association decisions. In particular, we propose to explore additional *context* information extracted from the user's devices that include the hardware (HW) type of the device (e.g., category and screen size) and the set of active applications. By becoming aware of the UEs' context and QoS requirements, the network can make more accurate decisions on *which* UE should be serviced, by which SBS, and *when*. Consequently, such a context-aware UE-SBS association scheme allows to provide precise QoS measures tailored to each UE's context and based on the actual requirements of each UE's applications.

The concept of context awareness, as complementary to location awareness, has been widely studied in pervasive computer science, and it is relatively novel in wireless networks [11]. Thereby, context information is typically used to recognize a network condition or the deployment scenario [11](and references therein), hence, for passive, monitoring, operations. In contrast, active network operations, such as most existing cell association schemes, particularly for small cell networks, have been restricted to physical layer aspects [2–6], [9]. In this respect, we define new, composite (i.e., multi-dimensional) QoS requirements for multi-tasking UEs (e.g., smartphones or tablets). By doing so, we are able to devise better-informed UE-SBS associations, which meet application and device-specific QoS requirements. To our best knowledge, beyond [12] which addressed context-awareness for scheduling in macro-cell networks, little work has been done to develop context-aware UE-SBS association schemes tailored to small cell networks.

The main contribution of this paper is to study and design novel strategies for context-aware SBS-UE association in the downlink of wireless small cell networks. We formulate the problem as a matching game in which the SBSs and the UEs are the players that need to rank one another so as to find a suitable association. The ranking is done by using utility functions that properly capture the application and device context information (at the UE side) as well as the interference and current network congestion (at the SBS side). The key advantage of the proposed model and approach lie in the fact that the UE-SBS association is achieved through distributed decisions at each UE and SBS that are based on practical UEs' context information such as the individual application set, their QoS needs, and the devices' hardware characteristic. We show that the performance of each UE and SBS is strongly affected by the dynamic formation of other UE-SBS links due to the dependence of the utility functions on externalities. For the proposed game, these externalities relate to the unique features of small cell networks such as interference limitations, spectrum availability, and network congestion. While some work on matching games in wireless networks exist such as in [7], [10], [13], these works are not tailored to small cell networks and do not account for externalities as done here. In fact, the works in [7], [10], [13] assume that the individual players' utilities are unaffected by the other player's preferences, which is impractical for the studied small cell association problem. To solve the proposed matching game with externalities, we propose a novel, distributed algorithm that allows the UEs and SBSs to interact so as to optimize their interdependent utilities. We show that the proposed algorithm efficiently handles the game's externalities and allows the SBSs and UEs to self-organize into a suitable stable UE-SBS matching. This stable outcome accounts for the available context at both the SBS and UE sides. Simulation results assess the various properties of the proposed approach and show significant performance gains compared to baseline context-unaware cell association.

The rest of this paper is organized as follows. In Section II, we present the system model and we introduce the concept of user context. In Section III, we formulate the UE-SBS association problem as a matching game with externalities, and we propose a novel algorithm to solve it. Simulation results are analyzed in Section IV. Finally, conclusions are drawn in Section V.

## II. SYSTEM MODEL

### A. Network Model

Consider the *downlink* transmission of a single orthogonal frequency division multiple access (OFDMA) macrocell (e.g., an LTE-Advanced or WiMAX network). In this network, $M$ UEs and $N$ SBSs are deployed. Let $\mathcal{M} = \{1,...,M\}$ and $\mathcal{N} = \{1,...,N\}$ denote, respectively, the set of all UEs and all SBSs. In conventional systems [8], each UE is typically serviced by the SBS with the highest receive signal strength indicator (RSSI). Here, we denote by $\mathcal{L}_i$ the set of UEs serviced by an SBS $i$ and, by $w_{i,m}$, the bandwidth that SBS $i$ allocates to each UE $m \in \mathcal{L}_i$. The transmit power for each transmission to an UE $m \in \mathcal{L}_i$ is denoted by $p_i$. The packet generation process at SBS $i$ is modeled as an M/D/1 queuing system. Here, the aggregated input traffic of UE $m \in \mathcal{L}_i$ is composed by packets of constant size generated using a Poisson arrival process with an average arrival rate of $\lambda_m$, in bits/s. For the transmission of these packets, the capacity between SBS $i$ and UE $m$ is given by:

$$\mu_{i,m}(\gamma_{i,m}) = w_{i,m} \log(1 + \gamma_{i,m}), \quad (1)$$

where $\gamma_{i,m} = \frac{p_i h_{i,m}}{\sigma^2 + I_{i,m}}$ is the signal-to-interference-plus-noise ratio (SINR) with $h_{i,m}$ indicating the channel gain between SBS $i$ and UE $m$ and $\sigma^2$ the variance of the Gaussian noise. Here, the interference component $I_{i,m} = \sum_{j \neq i} p_j h_{j,m}$, $j \in \mathcal{N} \setminus \{i\}$ relates to the transmissions from other SBSs $j$ to their respective UE $n \in \mathcal{L}_j$, which use the same subchannels of $w_{i,m}$. $p_j$, and $h_{j,m}$ respectively denote the transmit power and the channel realization between SBS $j$ and UE $m$.

The probability of packet error during the transmission between an SBS $i$ and a UE $m$, can be expressed via the probability of having the SINR below a target level $\Gamma_i$, and, for uncoded quadrature amplitude modulation (QAM)[1], this packet error rate (PER) is given by:

$$\text{PER}_{i,m}(\gamma_{i,m}) = \begin{cases} e_i \exp(-f_i \gamma_{i,m}), & \text{if } \gamma_{i,m} \geq \Gamma_i, \\ 1, & \text{otherwise,} \end{cases} \quad (2)$$

where $e_i, f_i$ are packet-size dependent constants and $\Gamma_i$ is a minimum SINR threshold for the correct demodulation. For ease of analysis, we do not consider the retransmission of the packets which are erroneously received. In such a conventional approach where the SBS has little practical information on the UE type, each SBS considers that the traffic streams of its UEs have the same priority, such as in [1]. Thus, it will schedule them with a uniform probability. In this respect, the delay for each UE $m \in \mathcal{L}_i$ depends on the aggregated input traffic of the other UEs $n \in \mathcal{L}_i \setminus \{m\}$, serviced by SBS $i$, which can be computed by combining the traffic arrival rates: $\lambda_{i,m} = \lambda_m + \sum_{n \in \mathcal{L}_i} \lambda_n$. Consequently, for a given UE $m$ served by SBS $i$, we express the average delay by Little's theorem [14]:

$$d_{i,m} = \frac{\lambda_{i,m}}{2\mu_{i,m}(\mu_{i,m} - \lambda_{i,m})}. \quad (3)$$

From the above formulation, we note that the performance of a UE $m$ serviced by SBS $i$ is affected by both the interference from the other noncooperative SBSs and the traffic generation process of the remaining UEs $n \neq m$ serviced by SBS $i$.

### B. Notion of UEs' Context

In existing networks, the best serving SBS is identified based on indicators of the wireless link quality at the UEs, such as the received signal strength indicators (RSSIs) or the SINRs [1]. Such existing UE-SBS association schemes are based solely on the physical properties of the wireless link, and, thus, they lead to two key drawbacks. First, the SBSs are unable to differentiate between individual traffic requests generated from each UE's application. Therefore, the traffic streams are typically scheduled independently with an uniform priority. However, in practical systems, not all applications have the same priority at the UE side. This is particularly important for modern smartphones which can host a diverse number of applications. For example, applications with "always active" traffic request (e.g., HD video streaming) or "keep alive"-type applications with periodic traffic, that mostly relate to feeds and notifications. Second, given the richness of emerging wireless services, features such as the *hardware* capabilities of the UE play a key role in the perceived QoS at the UE side. For example, for devices with large screen size and high resolution such as tablets and laptops, the user's QoS perception of video applications is more sensitive than on smaller devices such as smartphones. Due to the relevance of the set of active applications at the UE and its HW characteristics (e.g., the screen size is a key factor for determining the QoS of video services), we define the user's *context* as the set of all the relevant information that relates to the UE's hardware type and the properties of its active applications. Despite the important role of such context information in determining QoS provisions, existing approaches for cell association in small cell networks, such as [1–3], [6], [9], do not take that into account. In fact, the knowledge of the UEs' context information leads to new perspectives for the problem of SBS-UE association, by allowing to differentiate the UEs' QoS based on individual properties of the UEs' devices; hence yielding a smarter resource allocation.

---

[1]Nevertheless, the proposed solution can accommodate any other modulation scheme, without loss of generality.

TABLE I
TYPICAL QOS REQUIREMENTS OF MULTIMEDIA APPLICATIONS [1].

| Application | Data rate [kbps] | Delay [ms] | PER |
|---|---|---|---|
| HD video streaming | 800 | 2000 | 0.05 |
| Video conferencing | 700 | 30 | 0.01 |
| VoIP | 512 | 150 | 0.01 |
| Audio streaming | 320 | 200 | 0.08 |
| File download | 200 | 3000 | 0.1 |

In order to capture such context information, for each UE $m \in \mathcal{M}$, we construct an $a_m \times b_m$ dimension matrix $\mathbf{A}_m$ that reflects the practical QoS parameters of popular wireless services such as those shown in Table I (naturally, this matrix can also accommodate any other application). Here, $a_m$ represents the number of active applications and $b_m$ the number of minimum QoS requirements.

We model the UE hardware type by distinguishing three UE categories, depending on their screen size: smartphones, tablets, and laptops. These categories evenly partition the set of UEs $\mathcal{M}$. For the UE's application set, we propose that each UE $m$ constructs an $a_m \times 1$ dimension vector $\mathbf{g}_m$, with each component $g_{m,x}$ representing a *priority* for the $x$-th active application in $\mathbf{A}_m$. Such priorities are defined as follows. If $\mathbf{A}_m$ includes video applications, these will have the highest priority for tablets and laptops. File downloads are assigned the lowest priority, since they often run as background applications. For any other combination of the active applications, the priority is arbitrarily defined by the UE. Consequently, the context of each UE $m$ is defined by its active applications in $\mathbf{A}_m$ and their respective priorities in $\mathbf{g}_m$. For example in (4), we show an illustrative example of $\mathbf{A}_m$ and $g_m$ for a tablet UE $m$ with HD video streaming as the main application ($g_{m,2} = 1$), followed by VoIP $g_{m,1} = 2$ and file download ($g_{m,3} = 3$).

$$\mathbf{A}_m = \begin{bmatrix} 512 & 150 & 0.01 \\ 800 & 2000 & 0.05 \\ 200 & 3000 & 0.1 \end{bmatrix}, \mathbf{g}_m = \begin{bmatrix} 2 \\ 1 \\ 3 \end{bmatrix}. \quad (4)$$

Using this model, we are able to define a context-aware UE-SBS association scheme that differentiates and prioritizes the traffic generated from different applications and UE hardware types. In particular, for an aggregated traffic $\lambda_m$ of UE $m$, each SBS $i$ is able to discriminate the traffic stream of each application $\lambda_{m,x}$, for which $\sum_{x=1}^{a_m} \lambda_{m,x} = \lambda_{i,m}$. Based on this, each SBS is able to schedule each traffic stream $x$ with priority $k = g_{m,x}$, as extracted from UE $m$'s context. In this context-aware case, the traffic at each SBS is modeled as a priority-based M/D/1 queueing system. In such a system, the traffic requests of each UE $m$ are serviced according to the context-dependent priorities in $\mathbf{g}_m$. The delay of UE $m$ depends on the traffic load of the other UEs $n \in \mathcal{L}_i$ currently serviced by SBS $i$. Here, without loss of generality, we consider a nonpreemptive policy in which the traffic requests of a high priority user can move ahead of all the low priority traffic waiting in the queue. However, low priority packets in service are not interrupted by the higher priority users' packet arrivals. Thus, the UEs $n \in \mathcal{L}_i$ incur an initial delay for UE $m$ denoted by $D_m(\mathcal{L}_i)$. In this scenario, the average delay for the $k$-th priority stream of UE $m$ serviced by SBS $i$ is given by:

$$d_{i,m}^k = \frac{\sum_{x=1}^{a_m} \lambda_{m,x} \bar{M}_m^2}{2(1 - \sum_x^{k-1} \rho_{m,x})(1 - \sum_x^k \rho_{m,x})} + \frac{1}{\mu_{i,m}} + D_m(\mathcal{L}_i), \quad (5)$$

where $\rho_{m,x} = \frac{\lambda_{m,x}}{\mu_{i,m}}$ is the utilization factor for the $x$-th stream of UE $m$ and $\bar{M}_m^2$ the second moment of service time. By comparing the delay expressions in (3) and in (5), we can clearly see that the knowledge of context information enables each SBS to better prioritize application requests. In addition, context-aware SBSs and UEs are able to devise better-informed associations by guaranteeing the QoS constraints of each individual traffic requests.

With these considerations in mind, we propose that upon cell association, each UE and the SBSs in its vicinity exchange information on the UE's context and the SBSs' average performance metrics[2] $\mathbf{g}_m$. Note that such information exchange solely involves UEs and SBSs in the vicinity of one another.

## III. CELL ASSOCIATION AS A MATCHING GAME WITH EXTERNALITIES

### A. Problem Formulation

For associating UEs to SBSs, each SBS aims at identifying the largest set of UEs, for which it can meet the respective QoS requirements. In order to formalize the UE-SBS association problem, we define a suitable context-aware utility function for a UE $m \in \mathcal{L}_i$ serviced by SBS $i \in \mathcal{N}$ as follows:

$$U_{i,m}(\mathbf{A}_m, \mathbf{g}_m, \gamma_{i,m}, \eta) = \frac{\mu_{i,m}(\eta)(1 - \text{PER}_{i,m}(\eta))}{\sum_k^{a_m} d_{i,m}^k(\mathbf{g}_m, \eta)}. \quad (6)$$

Note that this utility function captures the data rate and packet error rate that SBS $i$ can deliver, given the SINR $\gamma_{i,m}$. Moreover, the utility in (6) properly accounts for the UE's required QoS and context in terms of applications (through $\mathbf{A}_m$) and hardware type (through $\mathbf{g}_m$).

Having defined such utility, we aim at solving the problem of assigning each UE $m \in \mathcal{M}$ to the best serving base station $i \in \mathcal{N}$ through a matching $\eta : \mathcal{M} \to \mathcal{N}$. Essentially, this yields the following optimization problem:

$$\arg\max_{\eta : (i,m) \in \eta} \sum_{i \in \mathcal{N}} \sum_{m \in \mathcal{L}_i} U_{i,m}(\mathbf{A}_m, \mathbf{g}_m, \gamma_{i,m}, \eta) \quad (7)$$

---
[2]This feedback can be done on traditional control channels such as in [1].

$$\text{s.t.,} \quad \mu_{i,m}(\gamma_{i,m},\eta) \geq \max_x \mathbf{A}_m(x,1), \forall m \in \mathcal{M}, i \in \mathcal{N} \quad (8)$$

$$d_{i,m}^k(\mathbf{g}_m,\eta) \leq \mathbf{A}_m(x,2), \forall k, \; k = g_{m,x}, \; \forall m \in \mathcal{M}, i \in \mathcal{N} \quad (9)$$

$$\text{PER}_{i,m}(\gamma_{i,m},\eta) \leq \min_x \mathbf{A}_m(x,3), \forall (i,m) \in \eta. \quad (10)$$

Note that the above optimization problem is subject to context dependent QoS constraints. Namely, constraint (8) ensures that each SBS-UE link $(i,m) \in \eta$ satisfies the most stringent requirements of data rate. Constraint (9) accounts for individual delay constraints of each of the applications in $\mathbf{A}_m$. Finally, constraint (10) captures the minimum requirement of packet error rate per application of each UE $m$.

In terms of complexity, solving the UE-SBS association using classical optimization techniques as per (7) is an NP-hard problem, which depends on the number of SBSs and UEs in the network. Even by relaxing some of the constraints in (8)-(10), the exponential complexity renders a centralized approach intractable, especially for small cell networks in which the number of UEs and SBSs can significantly grow. This complexity coupled with the need for self-organizing solutions in small cells mandate a distributed approach in which UEs and SBSs autonomously decide on the best UE-SBS association, based on their individual objectives. Accordingly, we propose that a distributed approach that accounts for the individual decisions and context information available at the UEs and SBSs, based on the context-aware utility as per (6). One suitable tool for developing such a self-organizing SBS-UE cell association approach which can solve (7) (while avoiding combinatorial complexity) is given by the framework of matching games [15]:

**Definition 1:** A *matching game* is defined by two sets of players $(\mathcal{M}, \mathcal{N})$ and two preference relations $\succ_m, \succ_i$ allowing each player $m \in \mathcal{M}, i \in \mathcal{N}$ to build preferences over one another, i.e., to rank, respectively, the players in $\mathcal{N}$ and $\mathcal{M}$.

The outcome of a matching game is a matching function (or association) $\eta$ that bilaterally assigns to each player $m \in \mathcal{M}$, a player $i = \eta(m), i \in \mathcal{M}$, and vice versa (i.e., $m = \eta(i)$). Here, a preference relation $\succ$ is defined as a complete, reflexive, and transitive binary relation between the players in $\mathcal{M}$ and $\mathcal{N}$. Thus, for any UE $m$, a preference relation $\succ_m$ is defined over the set of SBSs $\mathcal{N}$ such that, for any two SBSs $i,j \in \mathcal{N}^2, i \neq j$, and two matchings $\eta, \eta' \in \mathcal{M} \times \mathcal{N}, i = \eta(m), j = \eta'(m)$:

$$(i,\eta) \succ_m (j,\eta') \Leftrightarrow$$
$$U_{i,m}(\mathbf{A}_m, \mathbf{g}_m, \gamma_{i,m}, \eta) > U_{j,m}(\mathbf{A}_m, \mathbf{g}_m, \gamma_{j,m}, \eta'). \quad (11)$$

Similarly, for any SBS $i$ a preference relation $\succ_i$ over the set of UEs $\mathcal{M}$ is defined as follows, for any two UEs $m, n \in \mathcal{M}$, $m \neq n$ and two matchings $\eta, \eta' \in \mathcal{M} \times \mathcal{N}, m = \eta(i), n = \eta'(i)$:

$$(m,\eta) \succ_i (n,\eta') \Leftrightarrow$$
$$U_{i,m}(\mathbf{A}_m, \mathbf{g}_m, \gamma_{i,m}, \eta) > U_{i,n}(\mathbf{A}_n, \mathbf{g}_n, \gamma_{i,n}, \eta'). \quad (12)$$

By observing (11) and (12), we can see that the preferences of each UE over the set of SBS $\mathcal{N}$ depend on the existing matching $\eta$ in place in the network. In fact, for a UE-SBS link $(i,m) \in \eta$, the data rate in (1) and the PER in (2) depend on the interference produced by the other UE-SBSs links $(j,n) \in \eta$, $(i,m) \neq (j,n)$. Similarly, the delay of UE $m$ as per (5) is affected by the contexts of other users $n \in \mathcal{L}_i$ serviced by SBS $i$. As a result, for the studied problem, the preferences of UEs and SBSs are *interdependent*, i.e., they are influenced by the existing matching.

While most literature that deals with matching games, such as [7], [9], [10], [13], assumes that the preferences of a player do not depend on the other players choices, this assumption does not hold for the considered UE-SBS association problem. Such external effects that dynamically affect the performance of each UE-SBS link, are called externalities, and a suitable framework for studying them is given by *matching games with externalities* [15]. Unlike conventional matching games, when dealing with externalities, the potential matching $(i,m)$ between an SBS $i$ and a UE $m$ depends on the other UE-SBS associations in $\eta \setminus (i,m)$.

These externalities captured in the preferences in (11) and (12) lead to two important considerations. First, traditional concept solutions based on preference orders, such as the deferred acceptance algorithm used in [7], [10], [13], are unsuitable as the ranking of the preference changes as the matching forms. Second, choosing greedy utility-maximizing preferences does not ensure matching stability. In fact, due to externalities, a player may continuously change its preference order, in response to the formation of other UE-SBS links, and never reach a final UE-SBS association, unless externalities are well-handled.

### B. Proposed Solution and Algorithm

To solve the problem in (7) in a decentralized approach, we propose that the SBSs and UEs define individual preferences over one another, based on the preference relations in (11) and (12). The aim of each UE (SBS) is to maximize its own utility, or equivalently, to become associated with the most preferred SBS (UE)[3]. Due to the externalities, we look at

---

**Algorithm 1:** UE-SBS Cell Association Algorithm.

**Data**: Each UE is initially associated to a randomly selected SBS $i$.
**Result**: Convergence to a stable matching $\eta$.
**Phase I - SBS Discovery and Utility Computation**;
• Each UE $m$ discovers the SBSs in the vicinity;
• UEs and SBSs exchange context data $(\mathbf{A}_m, \mathbf{g}_m)$ and networks performance metrics in (1), (2), (5);
**Phase II - Swap-matching Evaluation**;
**repeat**
 • The utility $U_{i,m}(\eta)$ is updated based on the current $\eta$;
 • UEs and SBSs are sorted by $\succ_m$ and $\succ_i$;
 **if** $(j, \eta_{i,j}^m) \succ_m (i,\eta)$ **then**
  • Each UE $m$ sends a proposal to SBS $j$;
  • SBS $j$ computes $U_{j,m}(\eta_{i,j}^m)$ for the swap matching $\eta_{i,j}^m$;
  **if** $(m, \eta_{i,j}^m) \succ_j (m,\eta)$ **and** *(8)-(10) are satisfied* **then**
   • $\mathcal{L}_j \leftarrow \mathcal{L}_j \cup \{m\}$;
   • $\eta \leftarrow \eta_{i,j}^m$;
  **else**
   • SBS $j$ refuses the proposal, and UE $m$ sends a proposal to the next preference.
  **end**
 **end**
**until** $\nexists \eta_{i,j}^m : (j, \eta_{i,j}^m) \succ_m (i,\eta)$ *and* $(m, \eta_{i,j}^m) \succ_j (m,\eta)$;
**Phase III - Context-aware Resource Allocation**;
• For each active link, the SBSs initiate the context-aware transmissions as described in Section II.

---

[3]Henceforth, for notation simplicity, we only highlight the dependence of the utility in (6) on the current matching, while implying users' context information.

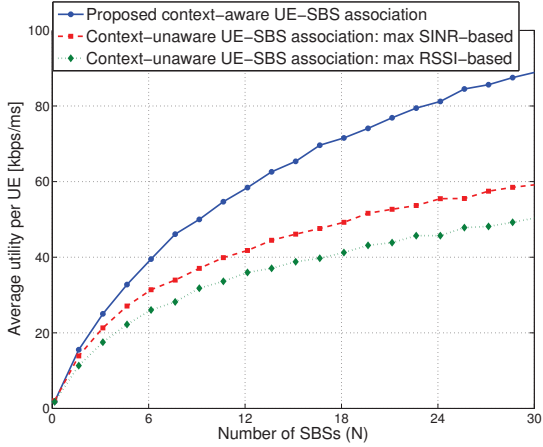
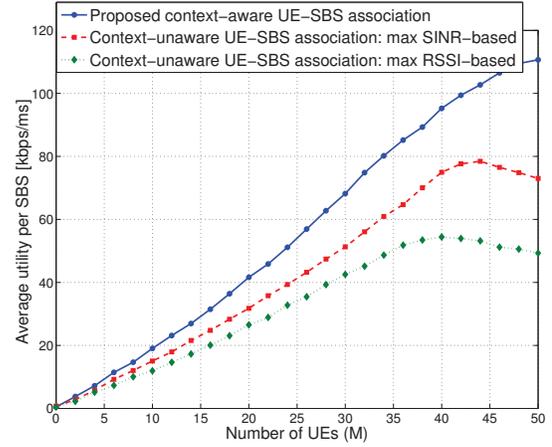

Fig. 1. Average individual utility per UE as a function of the number of SBSs $N$, under the considered approaches, $M = 60$ UEs, $a_m = 3$.

Fig. 2. Average individual utility per SBS as a function of the number of UEs $M$, $N = 25$ SBSs, $a_m = 3$.

a new stability concept, based on the idea of swap-matching [16]:

**Definition 2:** Given a matching $\eta$, a pair of UEs $m, n \in \mathcal{M}$ and SBSs $i, j \in \mathcal{N}$ with $(i, m), (j, n) \in \eta$, a swap-matching is defined as $\eta_{i,j}^m = \{\eta \setminus (i, m)\} \cup (j, m)$. Here, a matching is stable if there exist no swap-matchings $\eta_{i,j}^m$, such that:
- $\forall x \in \{m, n, i, j\}$, $U_{x, \eta_{i,j}^m(x)}(\eta) > U_{x, \eta(x)}(\eta)$ and
- $\exists x \in \{m, n, i, j\}$, $U_{x, \eta_{i,j}^m(x)}(\eta) \geq U_{x, \eta(x)}(\eta)$.

Given the notion of a stable swap-matching, at the network level, a matching $\eta$ with link $(i, m) \in \eta$ is said to be *stable* if there does not exist any UE $n$ or SBS $j$, for which SBS $i$ prefers UE $n$ over UE $m$, or any UE $m$ which prefers SBS $j$ over $i$. Such network-wide matching stability is reached by guaranteeing that swaps occur if they are beneficial for the involved players (i.e. $\{m, n, i, j\}$), given the externalities in the current matching $\eta$. In fact, in a swap-matching $\eta_{i,j}^m$, a UE $m$ can only switch from an SBS $i$ to an SBS $j$, if this strictly increases the utility for any of the players in $\{m, n, i, j\}$, without decreasing the utilities of the other players (both UEs and SBSs). Thus, through swap-matchings, the order of preferences for each player not involved in the swap is unaltered. Hence, no player has an incentive to swap from its current association, leading to a network-wide stable SBS-UE association.

To find a stable matching for the small cell user association problem in (7), we propose Algorithm 1, composed of three main phases: SBS discovery, swap-matching evaluation, and context-aware resource allocation. Initially, each UE is associated to a randomly[4] selected SBS $i$. Then, each UE $m$ discovers the SBSs $j \in \mathcal{N}$ in the vicinity, using standard techniques such as in [1]. Next, UE $m$ exchanges its context information (i.e., $\mathbf{A}_m$ and $\mathbf{g}_m$) with SBS $j$, which, in turn, informs the UE $m$ on its performance metrics $\mu_{i,m}(\gamma_{i,m}, \eta)$, $\text{PER}_{i,m}(\gamma_{i,m}, \eta)$) and $d_{i,m}^k(\mathbf{g}_m, \eta)$, based on the current matching $\eta$. In the second phase, based on the current matching, UEs and SBSs update their respective utilities and individual preferences over one another. If a UE $m$ is not currently served by its most preferred SBS (denoted by $j$), it sends SBS $j$ a matching proposal. Upon receiving a proposal, SBS $j$ updates its utility and accepts the request of the UE only if strictly beneficial in terms of utility $U_{j,m}(\eta_{i,j}^m)$. Otherwise, if

[4]Equivalently, the UE can be initially associated to the closest SBS or to the MBS.

rejected, UE $m$ proposes to the next SBSs in its preference list. Both UEs and SBSs periodically update their respective utilities and preferences according to the current matching and ensure that they are associated to their respective first preference. The convergence of Algorithm 1 follows from:

**Lemma 1:** Upon convergence of Phase II, Algorithm 1 reaches a stable matching.

*Proof:* The proof follows from two considerations. First, due to their transmission range, an UEs can only reach a limited number of SBSs in its vicinity, and thus, the number of possible swaps is *finite*. Moreover, only the swaps which strictly increase a player's utility can occur. Second, once all the possible swaps have been evaluated, Phase II terminates and each UE remains associated to the most preferred SBS, and vice versa. Therefore, no further improvement can be achieved by swaps among neighboring UEs and SBSs. ∎

## IV. SIMULATION RESULTS

For our simulations, we consider a single macro-cell with a radius of 1 km and a bandwidth of 20 MHz. In this cell, $M$ UEs and $N$ SBSs are uniformly deployed. The transmit power of each SBS $i$ is $p_i = 33$ dBm. Transmissions are affected by distance dependent path loss and shadowing according to 3GPP specifications [8]. The minimum SINR required by each UE is $\Gamma_i = 9.56$ dB [8], the noise level is $\sigma^2 = -121$ dBm. The traffic rates $\lambda_m$ are based on the typical values in Table 1 assuming a standard packet size of 2000 bytes [8].

For comparison purposes, we consider two additional schemes, which represent baseline solutions for the user cell association problem [1], [8]. In the first scheme, the UE is associated to the SBS providing the strongest RSSI, while, in the second scheme, each UE is assigned to SBS providing the strongest SINR. Naturally, in the above schemes, which are context-unaware, the UEs are scheduled with an even priority.

Figure 1 shows the average utility per UE as a function of the number of SBSs $N$, in a network with $M = 60$ UEs, using $a_m = 3$ applications. Figure 1 shows that, in the proposed context-aware approach, the UEs become associated to the SBSs which can jointly provide enhanced SINRs (and, thus, increased rates and successful transmission rates), and lower delays, via context-based associations. As a result, by making context-aware decisions, the resulting UE-SBS association significantly increases the UEs' utilities. For instance, Figure 1 shows that the proposed context-aware matching game yields significant performance gains, increasing with the network size $N$, reaching up to $48\%$ and $78\%$ relative to the maximum SINR-based and the RSSI-based criteria, respectively.

Figure 2 shows the average utility per SBS as a function of the number of UEs $M$, for a network with $N = 25$ SBSs, $a_m = 3$. Figure 2 shows that the SBS utility in the three studied schemes is ultimately limited by different factors. In context-unaware RSSI- and SINR-based schemes, each UE tends to become associated to the closest SBS. However, in such cases, the traffic load at each SBS can rapidly increase leading to larger delays and largely unequal loads. In contrast, the proposed context-aware approach is able to better balance the traffic load in the network, based on the knowledge of the UEs' context information (applications and HW type). As seen in Figure 2, the proposed context-aware approach yields an increased sum-utility while avoiding unequal traffic loads. For larger networks, the maximum achievable utility for all the studied schemes decreases due to the the increased interference, which grows with the number of SBS $N$, in the network. Nonetheless, Figure 2 shows that the proposed context-aware yields significant performance gains for all network sizes, reaching up to $52\%$ over the maximum SINR-based approach and $119\%$ over the RSSI-based approach, in a network with $M = 50$ UEs and $N = 25$ SBSs. As a result, from Figure 1 and Figure 2, we clearly see that the proposed approach brings relevant performance gains in terms of balanced traffic load distribution, increased rates, and reduced delays.

Figure 3 shows the average number of algorithm iterations (Phase II in Algorithm 1) needed by each player to achieve convergence to a stable matching, as a function of the number of SBSs in the network and the number of applications at each UE. In Figure 3, we can see that the number of applications $a_m$ affects the number of instances that each UE $m$ has to evaluate, due to the constraints in (7). For instance, the average number of algorithm iterations when each UE requires $a_m = 3$ applications is $14\%$ and $21\%$ larger than the cases with $a_m = 2$ and $a_m = 1$. However, even for SBSs network of reasonable size ($N \leq 50$), the average number of iterations never exceeds $4.1$. In summary, Figure 3 shows that the network converges to a stable matching by performing a reasonable number of algorithm iterations at each UE, hence requiring a low overhead.

## V. CONCLUSIONS

In this paper, we have presented a novel, context-aware approach to the cell association problem in small cell networks. Beyond including accurate QoS requirements, our cell association scheme accounts for the UEs' hardware type (e.g., smartphone, tablet, laptop), the application in use, and the UEs individual priorities over them. The proposed scheme brings forward the important advantage of providing the a UE-SBS association which can best accommodate the individual QoS requirements, based on the distributed knowledge of practical context information extracted from modern UE devices. We have formulated the problem as a matching game with externalities, in which the UEs and SBSs build preferences over one another so as to choose their preferred matching. In this game, the preferences are interdependent and are a function of the potentially resulting matching. We have proposed a context-based algorithm that enables both UEs and SBSs to generate a list of preferences, while accounting for the network externalities. We have shown that, with the proposed algorithm, SBSs and UEs reach a stable matching in a reasonable number of simulation iterations. Simulation results have shown that the proposed context-aware approach can provide significant gains in terms of increased data rates and reduced delays, reaching up to $52\%$, with respect to a traditional context-unaware SBS-UE association which is based on the maximum SINR.

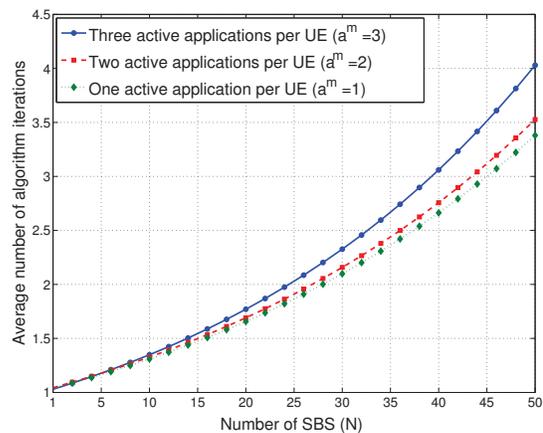

Fig. 3. Average number of iterations per UE for reaching a stable matching $\eta$, for different number of required applications $a_m$, $M = 50$ UEs.